\documentclass[aps,prl,twocolumn,showpacs,amsmath,amssymb]{revtex4-1}
\usepackage{pdfpages}

\usepackage[english]{babel}
\usepackage{amsmath}
\usepackage{bm}
\usepackage{graphicx} 
\usepackage{times}
\usepackage{epsfig} 
\usepackage[colorlinks,linkcolor=blue,citecolor=blue,urlcolor=blue]{hyperref}
\usepackage{color}
\definecolor{nred} {RGB}{224,0,0}
\definecolor{nblue} {RGB}{28,130,185}
\definecolor{dgreen} {RGB}{78,138,21}
\definecolor{norange}{RGB}{230,120,20}

\newcommand{\tr}{\mathrm{tr}}

\begin{document}

\title{Absence of full many-body localization in the disordered Hubbard chain}
\author{P. Prelov\v{s}ek$^{1,2}$}
\author{O. S. Bari\v si\'c$^3$}
\author{M. \v Znidari\v c$^2$}
\affiliation{$^1$~Jo\v zef Stefan Institute, SI-1000 Ljubljana, Slovenia}
\affiliation{$^2$~Faculty of Mathematics and Physics, University of Ljubljana, SI-1000 Ljubljana, 
Slovenia}
\affiliation{$^3$~Institute of Physics, Zagreb, Croatia}
\date{\today}
\begin{abstract}

We present numerical results within the one-dimensional disordered Hubbard model for several characteristic indicators of the many-body localization (MBL). 
Considering traditionally studied charge disorder (i.e., the same disorder strength for both spin orientations) we find that even at strong disorder all signatures 
consistently show that while charge degree of freedom is non-ergodic, the spin is delocalized and ergodic. This indicates the absence of the full MBL in 
the model that has been simulated in recent cold-atom experiments. Full localization can be restored if spin-dependent disorder is used instead.

\end{abstract}

\maketitle

{\it Introduction.--} 
The many-body localization (MBL) is a phenomenon whereby an interacting many-body system localizes due to disorder, 
proposed \cite{fleishman80,basko06} in analogy to the Anderson localization of noninteracting particles \cite{anderson58,mott68}. 
The MBL physics has attracted a broad attention of theoreticians. Yet, it has so far been predominantly
studied within the prototype model, i.e., the one-dimensional (1D) model of interacting spinless fermions with 
random potentials,
equivalent to the anisotropic spin-$1/2$  Heisenberg chain with random local fields.
Emerging from these studies are main hallmarks of the MBL state of the system: a) 
the Poisson many-body level statistics \cite{oganesyan07,torres15,luitz15,serbyn16,vasseur16}, 
in contrast to  the Wigner-Dyson one for normal ergodic systems, b) vanishing of d.c. transport at finite temperatures 
$T>0$,  including the $T \to \infty$ limit \cite{berkelbach10,barisic10,agarwal15,gopal15,lev15,steinigeweg15,barisic16,znidaric16},
c) logarithmic growth of the  entanglement entropy \cite{znidaric08,bardarson12,serbyn15},
as opposed to linear growth in generic systems, d) an existence of a set of local integrals of motion~\cite{huse14,serbyn13,Ros:2015,Imbrie:2016}, and e)
a non--ergodic time evolution of (all) correlation functions and of quenched initial states 
\cite{monthus10,pal10,luitz16,mierzejewski16,prelovsek16}. Because of these unique properties, the MBL can be used, e.g. 
to protect quantum information~\cite{huse13,Chandran:2014}. For more detailed review see 
Refs.~\cite{Nandkishore:2015,Altman:2015}.

The experimental evidence for the MBL comes from recent experiments on cold atoms
in optical lattices \cite{schreiber15,kondov15,bordia16,choi16} and ion traps~\cite{Maryland}. 
In particular, for strong disorders, experiments reveal non--ergodic decay 
of the initial density profile in uncoupled \cite{schreiber15} and coupled \cite{bordia16} 1D fermionic 
chains, as well as the vanishing of d.c. mobility in a 3D disordered lattice \cite{kondov15}.
In contrast to most numerical studies, being based on the spinless fermion models, the 
cold-atom experiments simulate a disordered  Hubbard model.  
The latter has been much less investigated theoretically
\cite{schreiber15,mondaini15,barlev16}, whereby results show that density imbalance might be non-ergodic 
at strong disorder \cite{schreiber15,mondaini15}, in accordance with experiments \cite{schreiber15,bordia16}.

The essential difference with respect to the interacting spinless model is that Hubbard model has two local degrees of 
freedom: charge (density) and spin. The aim of this Letter is to present numerical evidence that in the case of a
(charge) potential disorder and finite repulsion $U>0$ (as e.g. realized in the cold-atom experiments),
both degrees behave qualitatively different. In particular, while for strong disorder the charge 
exhibits non-ergodic behavior, e.g., the charge-density-wave and the local charge correlations fail to reach the thermal equilibrium, 
the spin imbalance and the local spin correlations show a clear decay. Similarly, we find that d.c. charge conductivity vanishes 
with the increasing disorder, whereas spin conductivity remains finite in the d.c. limit or is at least subdiffusive. 
The entanglement entropy, which incorporates both degrees, grows as a power law with time. 
All these findings reveal that even for strong disorders the system does not
follow the full MBL scenario, requiring the existence of a full set of local 
conserved quantities~\cite{Nandkishore:2015,huse14,serbyn13}. The present results point towards a novel phenomenon 
of a partial non-ergodicity and an effective dynamical charge-spin separation. Furthermore, we show that the 
localization of the spin degree of freedom may be achieved when the symmetry between the up and down fermions 
is lifted, for instance, by introducing a spin-dependent disorder.

{\it Model.--} 
The 1D disordered Hubbard model is given by the Hamiltonian,
\begin{equation}
H = - t_0 \sum_{js } ( c^\dagger_{j+1,s} c_{js} + \mathrm{h.c.}) + U \sum_j n_{j\uparrow}
n_{j\downarrow} + \sum_j \epsilon_j n_j\;. \label{hub}
\end{equation}
where $n_j=n_{j \uparrow} + n_{j \downarrow}$ is the local (charge) density. In our analysis, we consider the local (spin) magnetization as well, given by $m_j=n_{j \uparrow} - n_{j \downarrow}$. The quenched local potential disorder in Eq.~(\ref{hub}) involves a random uniform distribution $-W < \epsilon_j <W$. $t_0=1$ is used as the unit of energy.
In order to look for possible MBL features  of the whole many-body spectrum, we focus
our numerical calculations on the $T \to \infty$ limit.  With the average density $\bar n=\frac{1}{L}\sum_j n_j$ 
and the average magnetization $\bar m= \frac{1}{L}\sum_j m_j$ being constants of motion, we choose to investigate the unpolarized
system $\bar m=0$ and the half-filling $\bar n=1$ case, which is a generic choice at high $T$. Nevertheless, we also test the quarter-filling case, $\bar n=1/2$, see the Supplement~\cite{suppl}, as it is the one realized in experiments~\cite{schreiber15,bordia16}.

{\it Imbalance correlations.--}
In connection with cold-atom experiments are most relevant charge (density) imbalance correlations $I(t)$ as they evolve 
in time from an initial out-of-equilibrium configuration.  Therefore, we first discuss related 
charge-density-wave (CDW) and spin-density-wave (SDW) autocorrelation functions,
\begin{eqnarray}
C(\omega) = \frac{\alpha }{L } \mathrm{Re} \int_0^\infty  dt \mathrm{e}^{i \omega t} \langle n_{\pi}(t) n_{\pi} \rangle, 
\nonumber \\
S(\omega) = \frac{\alpha }{L } \mathrm{Re} \int_0^\infty  dt \mathrm{e}^{i \omega t} \langle m_{\pi}(t) m_{\pi} \rangle\;,\label{eq02}
\end{eqnarray}
calculated for a particular (staggered) wavevector $q=\pi$, with $n_{q=\pi}= \sum_j (-1)^{j} n_j$ for the CDW case, and 
$m_{q=\pi}= \sum_j (-1)^{j} m_j$ for the SDW case. In Eq.~(\ref{eq02}), 
$1/\alpha = \bar n(1-\bar n/2)$  so that  $C(t=0)=S(t=0)=1$, for $T, L \to \infty$. The non--ergodicity (after taking $L \to \infty$) should 
manifest itself as a singular contribution, $C(\omega \sim 0) =C_0 \delta(\omega)$, $S(\omega \sim 0)=S_0 \delta(\omega)$, 
with $C_0$ and $S_0$ corresponding to the CDW and the SDW stiffnesses, respectively. That is, the (full) MBL requires that both, 
$C_0$ and $S_0$,  are finite. For the calculation of imbalance correlations we employ 
the microcanonical Lanczos method (MCLM) \cite{long03,prelovsek13} on finite systems of maximum 
length  $L=14$ for $\bar n=1$  (for $\bar n=1/2$ see the Supplement~\cite{suppl} ). The high frequency resolution is achieved by large number of Lanczos steps $N_L = 10^4$, 
$\delta \omega \propto 1/N_L$.   The averaging over disorder realizations is performed over $N_s = 20 -100$ 
different $\epsilon_j$ configurations.

Instead of plotting spectra $C(\omega),S(\omega)$, given by Eq.~(\ref{eq02}), it is more informative to display quasi-time evolution  
$C,S(\tau)= \int_{-1/\tau}^{1/\tau} d\omega ~C,S (\omega)$. In this way we omit fast oscillations with typical
$\omega= t_0$,  while retaining the physical content of the limit $t=\tau \to \infty$. 
In Fig.~1 we compare results for $C(\tau)$ and $S(\tau)$ at half-filling $\bar n=1$ for intermediate $U=4$
and a wide span of  disorder $W = 2 - 15$, obtained by the MCLM for $L=14$. 
(In the Supplement \cite{suppl} we compare results obtained for different $L$, showing that they 
are mutually consistent for $L \geq 10$.)  Results are plotted up to maximum times $\tau_m =1/\delta \omega$, 
where for different $L\leq14$ we get $\tau_m =50 -200$, depending  on $W$. 

Results presented in Fig.~1 reveal qualitative difference between charge and
spin dynamics within the Hubbard model. For $C(\tau)$ we observe a behavior that is 
qualitatively very similar to the behavior of the density imbalance in the spinless model \cite{luitz15,mierzejewski16}, or to the behavior reported in experiments \cite{schreiber15,bordia16}.  Namely, in the presence of finite $U>0$, the CDW
correlations are ergodic $C(\tau \to \infty) \to 0$ for weak disorders 
$W =2,3$,  while for large disorders, e.g.  $W = 6, 15,$ the non--ergodicity appears, $C(\tau \to \infty) = C_0 >0$.
This is in clear contrast with the spin imbalance case, $S(\tau)$, which decays to zero even for the 
strongest disorder $W =15$. Although the ergodic-nonergodic transition from CDW correlations in Fig.~1 cannot be precisely located, $W^* \sim 4-6$, it is clearly there. On the other hand, no such transition can be observed in SDW correlations, which remain ergodic independently of disorder strength. 

\begin{figure}[!tb]
\includegraphics[width=0.9\columnwidth]{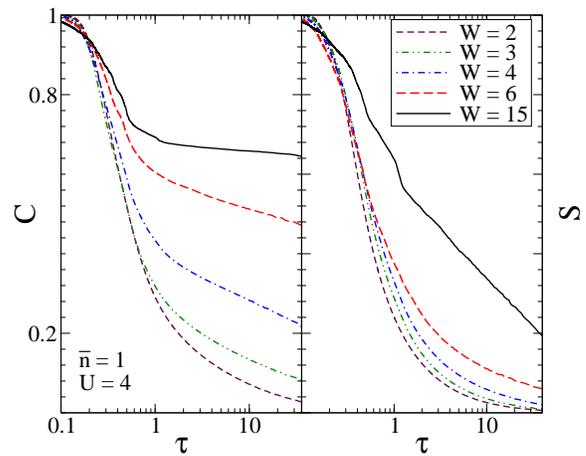}
\caption{(Color online) Charge and spin imbalance correlations $C(\tau)$ and $S(\tau)$,
respectively, as evaluated by the MCLM  at half--filling $\bar n=1$ and $U=4$, at fixed system size $L=14$.
The potential disorder is varied in the range $W=2 - 15$.}
\label{fig1}
\end{figure}
\begin{figure}[!tb]
\includegraphics[width=0.9\columnwidth]{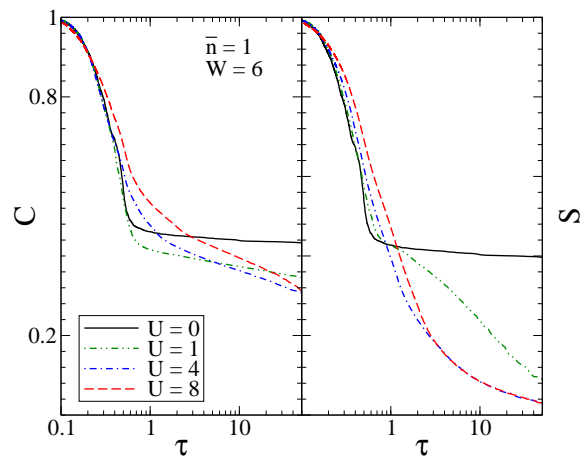}
\caption{(Color online) $C(\tau)$ and $S(\tau)$ calculated for half--filling $\bar n=1$ and $L=12$, for fixed disorder $W=6$
and  various interaction strengths $U=0 - 8$. }
\label{fig2}
\end{figure}

A similar message is obtained from $C,S(\tau)$, being presented in Fig.~2 for fixed $W$ as a function of interaction $U$.
In Fig.~2, the disorder strength is set to $W=6$, because for $U=4$ such $W$ corresponds to the non-ergodic regime for CDW correlations, as shown in Fig.~1. The noninteracting $U=0$ case is a particular one, 
involving the Anderson localization of single--particle states. Consequently, for $U=0$ both
$C(\tau)$ and $S(\tau)$ in Fig.~2 saturate to a constant value after a short transient $\tau \sim 1$.
For $U>0$, the behavior of $C(\tau)$ and $S(\tau)$ turns out to be very different. $C(\tau)$
exhibits a weak variation with $U >0$, but still with weak logarithmic-like time dependence \cite{mierzejewski16}. 
On the other hand, already the $U=1$ case leads to a decay of spin imbalance $S(\tau \to \infty)=0$. This decay becomes even faster for $U=4,8$.

{\it Local correlations.--}
Next we study local charge and spin dynamics, by considering the local real-time correlation $C_l(t)=A \sum_j \langle \rho_j(t) \rho_j \rangle $ and $S_l(t)=B \sum_j\langle m_j(t) m_j \rangle $, where $\rho_j=n_j-\bar n$, while $A$ and $B$ are normalization constants such that $C_l(0)=S_l(0)=1$. Similarly as for the imbalance, in a MBL system these two quantities freeze at a nonzero value~\cite{luitz16}, indicating the non-ergodicity. The advantage of the autocorrelation functions $C_l$ and $S_l$ over imbalance is that they exhibit smaller fluctuations for generic initial states.

\begin{figure}[!t]
\centerline{\includegraphics[width=0.9\columnwidth]{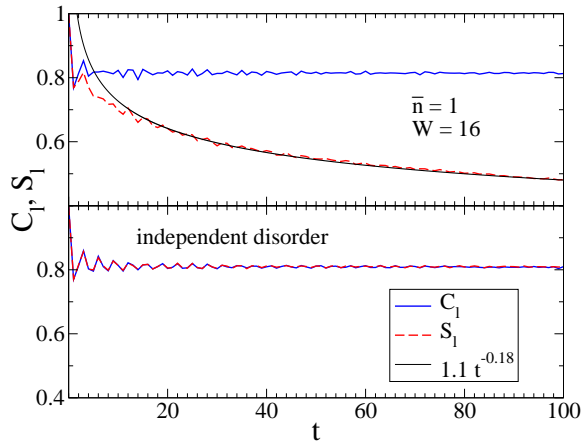}}
\caption{(Color online) Decay of the local charge and spin correlations for $U=1$ and $W=16$. (a) For the charge disorder, spin is delocalized (the red dashed curve). (b) For the independent disorder for each spin, the charge and the spin are both localized (note the two, red and the blue curve, almost completely overlapping). The averaging involves over $400$ product initial states, $L=64$.}
\label{fig:fluct}
\end{figure}

For the current analysis of the local correlations (as well as for calculations of the entropy afterwards), we use the time-dependent density matrix renormalization group method, which is an efficient method for evolution of initial product states provided the entanglement is small. For strong disorder we are typically able to simulate significantly larger systems ($L \approx 64$) than with the MCLM. Details of the method as well as references to original literature may be found in e.g. Ref.~\cite{znidaric08}. In Fig.~\ref{fig:fluct} we show results of such a simulation. One may see that even for very strong disorders $W$ and small interactions $U$ the spin autocorrelation function decays algebraically (unlike charge), again signaling the ergodicity of the spin degree of freedom. 

On the other hand, by considering a modification of the disorder model in Eq.~(\ref{hub}) and taking an independent disorder for the each spin orientation, i.e., $\sum_j (p_j n_{j\uparrow}+q_jn_{j\downarrow})$ with independent $p_j$ and $q_j\in [-W,W]$, a dramatic change occurs. As may be seen from Fig.~\ref{fig:fluct}b, now both, the spin and the charge, behave in the same way, freezing at a nonzero value, as expected for the MBL system.

{\it Dynamical conductivities.}
The question of d.c. transport is frequently analyzed in the context of 
dynamical charge and spin conductivities (or diffusivities, since we omit the prefactor $1/T$). In the $T \to \infty$ limit, these two conductivities are given by
\begin{equation}
\sigma_{c,s}(\omega) = \frac{1}{L}\mathrm{Re} \int_0^\infty  dt \mathrm{e}^{i \omega t} \langle j_{c,s}(t) j_{c,s} \rangle,
\end{equation}
where $j_{c,s}$ are charge and spin uniform currents, respectively, $j_{c,s} =  i \sum_{is } ( \pm 1)^s( c^\dagger_{i+1,s} c_{is} -  c^\dagger_{is} c_{i+i,s})$.

For the evaluation of $\sigma_{c,s}(\omega)$ we again employ the MCLM, using periodic boundary
conditions.  The numerical requirements are similar as for $C,S(\omega)$. Namely, the crucial role is played again by the high $\omega$
resolution, because the quantities of interest here are the d.c. value $\sigma_{c,s}(\omega \to 0)$ and the low--$\omega$ scaling of $\sigma_{c,s} (\omega)- \sigma_{c,s}(0)$ with $\omega$. 

\begin{figure}[!htb]
\includegraphics[width=0.9\columnwidth]{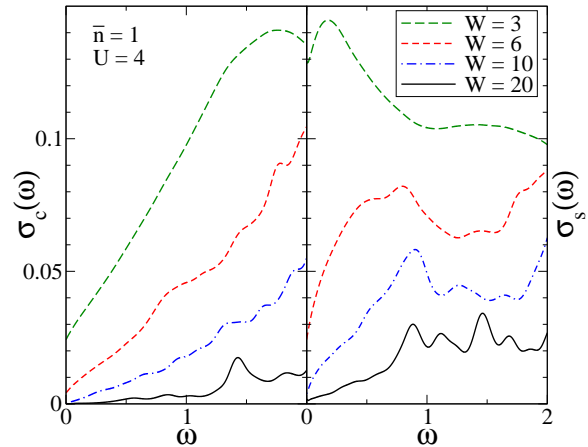}
\caption{(Color online) Charge and spin dynamical conductivity $\sigma_c(\omega)$ and 
$\sigma_s(\omega)$, respectively, evaluated at half--filling $\bar n=1$, $U=4$ at fixed size $L=14$,
but for various disorders $W=3 - 20$.}
\label{fig4}
\end{figure}

Results for $\sigma_c(\omega), \sigma_s(\omega)$ are presented in Fig.~4, for intermediate
$U=4$ and a wide range of disorders, $W=3 - 20$. It should be pointed out that due to insufficient sampling, $N_s$, the current results for strongest 
$W >10$ suffer in part from sample-to-sample fluctuations, which increase with $W$. On the other hand, the results for weaker $W$ are much less sensitive to fluctuations
 \cite{barisic16}. Conclusions that follow from $\sigma_c(\omega)$ in Fig.~4 are quite similar to those obtained for the spin-less model \cite{karahalios09,barisic10,steinigeweg15,barisic16}. 
 The maximum of $\sigma_c(\omega)$ at moderate disorder $W\geq 2$ is at $\omega_c^* \sim 2$, reflecting
the noninteracting limit. At low $\omega \ll1$, we find rather generic nonanalytical behavior 
$\sigma_c(\omega) \sim \sigma_c(0)+ \zeta |\omega|^\alpha$ with $\alpha \sim 1$. D.c. value $\sigma_s(0)$ is 
rapidly vanishing for $W> 4$. 

On the other hand, in Fig.~4, $\sigma_s(\omega)$ behaves qualitatively differently. 
In general, it exhibits two maxima, whereby the lower one at $\omega^*_s < 1$ is not present in $\sigma_c(\omega)$, indicating 
a different scale for the spin dynamics. In addition, finite $\sigma_s(0) > 0$ seems to be well resolved all the way up to $W = 20$.
Moreover, the low-$\omega$ behavior appears to be given by $\sigma_s(\omega) \sim \sigma_s(0)+ 
\xi |\omega|^\gamma$, with $\gamma <1$ even for the largest $W$. The implication of $\gamma<1$, being an indication 
of a subdiffusive dynamics \cite{agarwal15,agarwal115}, is divergent static magnetic polarizability 
$\chi_s \propto \int d\omega \sigma_s(\omega)/\omega^2$, 
even in the case of vanishing d.c. $\sigma_s(0)=0$. This low-frequency behavior of $\sigma_s(\omega)$ is compatible 
with a subdiffusive spin transport $\Delta m \sim t^{0.3}$, observed for 
initial states with global spin imbalance (see the Supplement~\cite{suppl} for details). Thus, spin (magnetization) is transported globally even for strong disorder.

\begin{figure}[ttp]
\centerline{\includegraphics[width=3.1in]{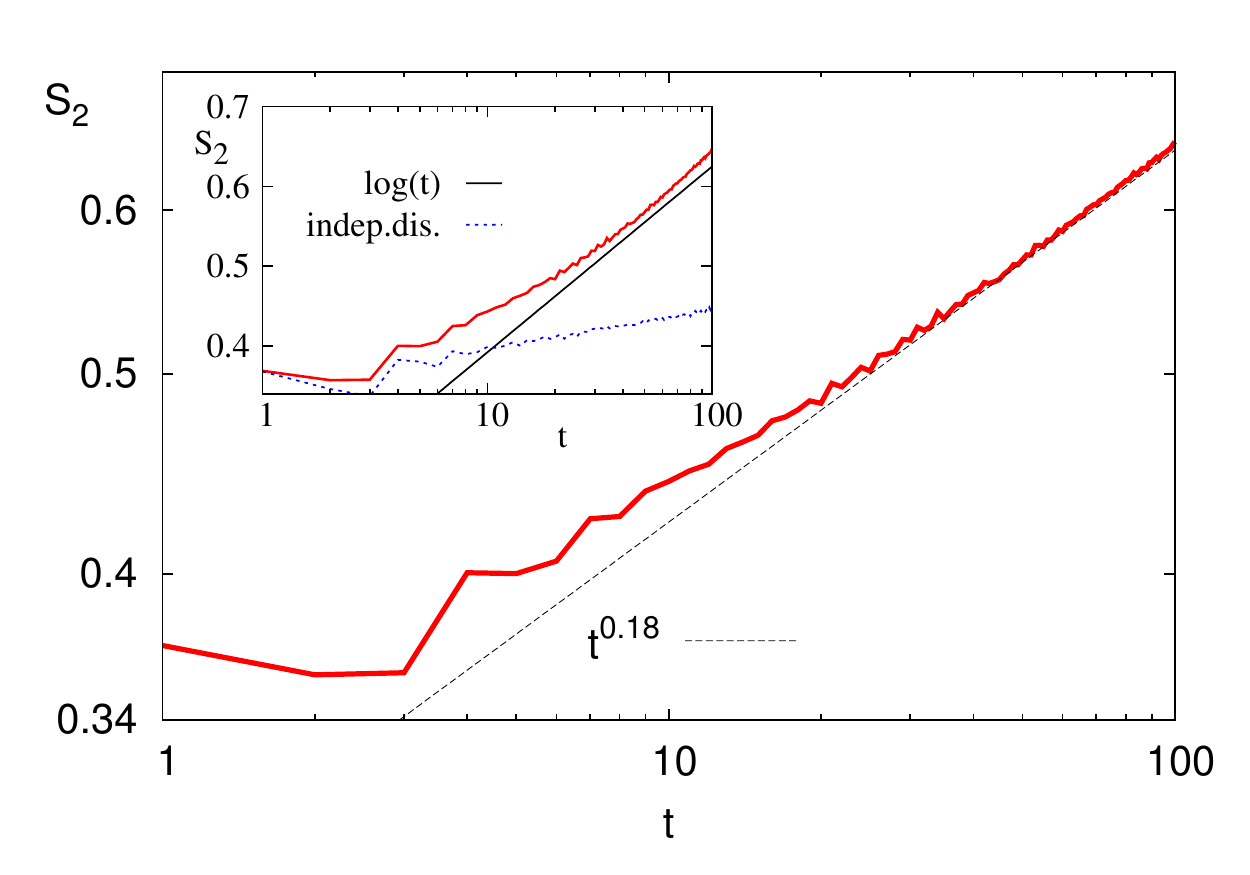}}
\caption{(Color online) Average von Neumann entropy $S_2(t)$ for $U=1$ and $W=16$ in a log-log plot. Inset: semi-log plot of the same data (red curve). For charge disorder (red curve) $S_2(t)$ is consistent with a power law, while for an independent disorder (blue curve in the inset) with a logarithmic growth. The same dataset as in Fig.~\ref{fig:fluct}, statistical fluctuations are of the size of curves thickness.}
\label{fig:S}
\end{figure}

{\it Entanglement entropy.--}
One of the defining properties of the MBL is logarithmic growth of entanglement with time~\cite{znidaric08}, when starting from a product initial state. In Fig.~\ref{fig:S} behavior of the entanglement entropy $S_2(t)=-\tr{[\rho_{\rm A}(t) \log_2\rho_{\rm A}(t)]}$ of the reduced density matrix $\rho_{\rm A}(t)$ is shown for $U=1$ and large $W$. From the semi-log plot (the inset in Fig.~\ref{fig:S}) one may see that $S_2(t)$ has a slight upward curvature, not growing logarithmically. Indeed, as show in main frame, the growth is better described by a power law, $S_2(t) \sim t^{0.18}$ (the power $\approx 0.18$ seems to be the same as the power of the decay of $S_l(t)$ in Fig.~\ref{fig:fluct}). On the other hand, with the independent disorder $W=16$ on both spin orientations one gets $S_2(t) \sim \log(t)$ (the blue curve in the inset in Fig.~\ref{fig:S}).

{\em Symmetry argument.--}
The ergodicity of the spin degrees has been so far established numerically. However, we wish to present additional symmetry arguments for $\bar n=1$ and $\bar m=0$ to demonstrate that $S_l(t \to \infty) \to 0$ for any fixed $L$. That is, in the absence of degeneracy, $S_l(t\rightarrow\infty)$ in the eigenbasis of $H$ is given solely by diagonal matrix elements of $m_j$. For charge disorder, $H$ is even under operation $P$ that exchanges up and down fermions. Consequently, all eigenstates for $U \neq 0$ have a well defined parity $P$, while $m_j$ is odd under $P$, and therefore all diagonal matrix elements of $m_j$ are zero by symmetry. 
The order of limits $L \to \infty, t \to \infty$ used above is opposite to the one required for a proof of ergodicity. Namely, there is always a possibility for the existence of an intermediate ``freezing'' timescale $t_f(L)$ at which $S_l(t_f)>0$, with $t_f(L)$ diverging in the thermodynamic $L\rightarrow\infty$ limit. However, our numerical data (see also Ref.~\cite{suppl}) does not give any hints for such behavior of $t_f(L)$.

{\it Conclusions.--}
We have presented numerical results for the 1D Hubbard model with random potentials, showing that the interacting fermion system does not exhibit the full MBL  up to very strong disorder, $W \leq 20$. Several indicators are inconsistent with accepted requirement for the MBL: 
a) spin imbalance correlations $S(t)$ decay to zero as
in ergodic systems, b) local spin correlations $S_l(t)$ decay to zero as well, although with a slow power-law decay, c) dynamical spin and charge conductivity behaves differently, i.e., we find finite d.c. value $\sigma_s(0) >0$, or at least subdiffusive $\sigma_s(\omega \to 0)$, for disorder strengths much above those for which $\sigma_c(0)$ vanishes, d) the entanglement entropy $S_2(t)$ does not saturate or increase logarithmically with $t$, but rather grows according to power law. While above findings rule out the existence of the full MBL in the model considered, they offer
a novel phenomenon which may be interpreted as a disorder induced dynamical charge-spin separation at all energy scales.
It should be pointed out that in a 1D disordered Hubbard model an effective charge-spin separation appears already 
at weak to modest $U \sim t_0$,  which should be distinguished from the $U \gg t_0$  limiting behaviors well known in pure model \cite{ogata90} and recently  reported also for a disordered model \cite{mondaini15,parameswaran16}. We cannot, however, exclude the possibility
that charge also would become ergodic at some very long time scale, which is so far beyond numerical  as well as experimental reach.

One might speculate that a particular absence of full MBL can be related to SU(2) symmetry  \cite{vasseur15,vasseur16,potter16} of the Hubbard model. 
Yet, the non-Abelian SU(2) spin rotation symmetry can be lifted by introducing a  constant-magnetic-field term $H'=B \sum_j m_j$, not changing our conclusions. Namely, time evolution of any state with a fixed number of up and down fermions remains the same. Therefore the presence or the absence of SU(2) symmetry is irrelevant for $T \to \infty$ averages (where all states have an equal weight) or for time evolution of specific states from any invariant subspace. It is also evident from our results that the above effective decoupling of charge and spin can be broken by e.g. an addition of random local (magnetic) fields. If fermions with different spin orientations exhibit independent disorder charge and spin can be both non-ergodic and one can have (full) MBL. There is also an interesting possibility that, if we use a spin disorder, i.e., $\sum_j \varepsilon_j (n_{j\uparrow}-n_{j\downarrow})$ instead of the charge disorder, the spin would be localized and the charge delocalized. Therefore, by a simple choice of disorder type we can tune transport properties of spin and charge -- a potentially useful property for engineered quantum devices. 
 
Our findings are not in disagreement with measurements of charge degree of freedom in cold-atom experiments, 
which simulate quarter-filled 1D Hubbard model and reveal a non-ergodic charge imbalance at 
strong quasi-periodic potential. We show in the Supplement~\cite{suppl} that with a random potential of similar strength the charge 
is non-ergodic, whereas spin correlations decay to zero, exhibiting no localization.

{\it Acknowledgments.}
P.P. acknowledges fruitful discussions with F. Pollmann and F. Heidrich-Meisner. 
P.P. and M. \v Z. acknowledge the support by the program P1-0044 and grant No. J1-7279 of the Slovenian Research Agency. O.S.B. 
acknowledge the support by the Croatian QuantiXLie Center of Excellence.

\bibliography{ref_manuhubmbl}

\begin{thebibliography}{52}%
\makeatletter
\providecommand \@ifxundefined [1]{%
 \@ifx{#1\undefined}
}%
\providecommand \@ifnum [1]{%
 \ifnum #1\expandafter \@firstoftwo
 \else \expandafter \@secondoftwo
 \fi
}%
\providecommand \@ifx [1]{%
 \ifx #1\expandafter \@firstoftwo
 \else \expandafter \@secondoftwo
 \fi
}%
\providecommand \natexlab [1]{#1}%
\providecommand \enquote  [1]{``#1''}%
\providecommand \bibnamefont  [1]{#1}%
\providecommand \bibfnamefont [1]{#1}%
\providecommand \citenamefont [1]{#1}%
\providecommand \href@noop [0]{\@secondoftwo}%
\providecommand \href [0]{\begingroup \@sanitize@url \@href}%
\providecommand \@href[1]{\@@startlink{#1}\@@href}%
\providecommand \@@href[1]{\endgroup#1\@@endlink}%
\providecommand \@sanitize@url [0]{\catcode `\\12\catcode `\$12\catcode
  `\&12\catcode `\#12\catcode `\^12\catcode `\_12\catcode `\%12\relax}%
\providecommand \@@startlink[1]{}%
\providecommand \@@endlink[0]{}%
\providecommand \url  [0]{\begingroup\@sanitize@url \@url }%
\providecommand \@url [1]{\endgroup\@href {#1}{\urlprefix }}%
\providecommand \urlprefix  [0]{URL }%
\providecommand \Eprint [0]{\href }%
\providecommand \doibase [0]{http://dx.doi.org/}%
\providecommand \selectlanguage [0]{\@gobble}%
\providecommand \bibinfo  [0]{\@secondoftwo}%
\providecommand \bibfield  [0]{\@secondoftwo}%
\providecommand \translation [1]{[#1]}%
\providecommand \BibitemOpen [0]{}%
\providecommand \bibitemStop [0]{}%
\providecommand \bibitemNoStop [0]{.\EOS\space}%
\providecommand \EOS [0]{\spacefactor3000\relax}%
\providecommand \BibitemShut  [1]{\csname bibitem#1\endcsname}%
\let\auto@bib@innerbib\@empty
\bibitem [{\citenamefont {Fleishman}\ and\ \citenamefont
  {Anderson}(1980)}]{fleishman80}%
  \BibitemOpen
  \bibfield  {author} {\bibinfo {author} {\bibfnamefont {L.}~\bibnamefont
  {Fleishman}}\ and\ \bibinfo {author} {\bibfnamefont {P.~W.}\ \bibnamefont
  {Anderson}},\ }\href {\doibase 10.1103/PhysRevB.21.2366} {\bibfield
  {journal} {\bibinfo  {journal} {Phys. Rev. B}\ }\textbf {\bibinfo {volume}
  {21}},\ \bibinfo {pages} {2366} (\bibinfo {year} {1980})}\BibitemShut
  {NoStop}%
\bibitem [{\citenamefont {Basko}\ \emph {et~al.}(2006)\citenamefont {Basko},
  \citenamefont {Aleiner},\ and\ \citenamefont {Altshuler}}]{basko06}%
  \BibitemOpen
  \bibfield  {author} {\bibinfo {author} {\bibfnamefont {D.}~\bibnamefont
  {Basko}}, \bibinfo {author} {\bibfnamefont {I.}~\bibnamefont {Aleiner}}, \
  and\ \bibinfo {author} {\bibfnamefont {B.}~\bibnamefont {Altshuler}},\ }\href
  {\doibase 10.1016/j.aop.2005.11.014} {\bibfield  {journal} {\bibinfo
  {journal} {Ann. Phys.}\ }\textbf {\bibinfo {volume} {321}},\ \bibinfo {pages}
  {1126} (\bibinfo {year} {2006})}\BibitemShut {NoStop}%
\bibitem [{\citenamefont {Anderson}(1958)}]{anderson58}%
  \BibitemOpen
  \bibfield  {author} {\bibinfo {author} {\bibfnamefont {P.~W.}\ \bibnamefont
  {Anderson}},\ }\href {\doibase 10.1103/PhysRev.109.1492} {\bibfield
  {journal} {\bibinfo  {journal} {Phys. Rev.}\ }\textbf {\bibinfo {volume}
  {109}},\ \bibinfo {pages} {1492} (\bibinfo {year} {1958})}\BibitemShut
  {NoStop}%
\bibitem [{\citenamefont {Mott}(1968)}]{mott68}%
  \BibitemOpen
  \bibfield  {author} {\bibinfo {author} {\bibfnamefont {N.~F.}\ \bibnamefont
  {Mott}},\ }\href {\doibase 10.1080/14786436808223200} {\bibfield  {journal}
  {\bibinfo  {journal} {Phil. Mag.}\ }\textbf {\bibinfo {volume} {17}},\
  \bibinfo {pages} {1259} (\bibinfo {year} {1968})}\BibitemShut {NoStop}%
\bibitem [{\citenamefont {Oganesyan}\ and\ \citenamefont
  {Huse}(2007)}]{oganesyan07}%
  \BibitemOpen
  \bibfield  {author} {\bibinfo {author} {\bibfnamefont {V.}~\bibnamefont
  {Oganesyan}}\ and\ \bibinfo {author} {\bibfnamefont {D.~A.}\ \bibnamefont
  {Huse}},\ }\href {\doibase 10.1103/PhysRevB.75.155111} {\bibfield  {journal}
  {\bibinfo  {journal} {Phys. Rev. B}\ }\textbf {\bibinfo {volume} {75}},\
  \bibinfo {pages} {155111} (\bibinfo {year} {2007})}\BibitemShut {NoStop}%
\bibitem [{\citenamefont {Torres-Herrera}\ and\ \citenamefont
  {Santos}(2015)}]{torres15}%
  \BibitemOpen
  \bibfield  {author} {\bibinfo {author} {\bibfnamefont {E.~J.}\ \bibnamefont
  {Torres-Herrera}}\ and\ \bibinfo {author} {\bibfnamefont {L.~F.}\
  \bibnamefont {Santos}},\ }\href {\doibase 10.1103/PhysRevB.92.014208}
  {\bibfield  {journal} {\bibinfo  {journal} {Phys. Rev. B}\ }\textbf {\bibinfo
  {volume} {92}},\ \bibinfo {pages} {014208} (\bibinfo {year}
  {2015})}\BibitemShut {NoStop}%
\bibitem [{\citenamefont {Luitz}\ \emph {et~al.}(2015)\citenamefont {Luitz},
  \citenamefont {Laflorencie},\ and\ \citenamefont {Alet}}]{luitz15}%
  \BibitemOpen
  \bibfield  {author} {\bibinfo {author} {\bibfnamefont {D.~J.}\ \bibnamefont
  {Luitz}}, \bibinfo {author} {\bibfnamefont {N.}~\bibnamefont {Laflorencie}},
  \ and\ \bibinfo {author} {\bibfnamefont {F.}~\bibnamefont {Alet}},\ }\href
  {\doibase 10.1103/PhysRevB.91.081103} {\bibfield  {journal} {\bibinfo
  {journal} {Phys. Rev. B}\ }\textbf {\bibinfo {volume} {91}},\ \bibinfo
  {pages} {081103} (\bibinfo {year} {2015})}\BibitemShut {NoStop}%
\bibitem [{\citenamefont {Serbyn}\ and\ \citenamefont
  {Moore}(2016)}]{serbyn16}%
  \BibitemOpen
  \bibfield  {author} {\bibinfo {author} {\bibfnamefont {M.}~\bibnamefont
  {Serbyn}}\ and\ \bibinfo {author} {\bibfnamefont {J.~E.}\ \bibnamefont
  {Moore}},\ }\href {\doibase 10.1103/PhysRevB.93.041424} {\bibfield  {journal}
  {\bibinfo  {journal} {Phys. Rev. B}\ }\textbf {\bibinfo {volume} {93}},\
  \bibinfo {pages} {041424(R)} (\bibinfo {year} {2016})}\BibitemShut {NoStop}%
\bibitem [{\citenamefont {Vasseur}\ \emph {et~al.}(2016)\citenamefont
  {Vasseur}, \citenamefont {Friedman}, \citenamefont {Parameswaran},\ and\
  \citenamefont {Potter}}]{vasseur16}%
  \BibitemOpen
  \bibfield  {author} {\bibinfo {author} {\bibfnamefont {R.}~\bibnamefont
  {Vasseur}}, \bibinfo {author} {\bibfnamefont {A.~J.}\ \bibnamefont
  {Friedman}}, \bibinfo {author} {\bibfnamefont {S.~A.}\ \bibnamefont
  {Parameswaran}}, \ and\ \bibinfo {author} {\bibfnamefont {A.~C.}\
  \bibnamefont {Potter}},\ }\href {\doibase
  http://dx.doi.org/10.1103/PhysRevB.93.134207} {\bibfield  {journal} {\bibinfo
   {journal} {Phys. Rev. B}\ }\textbf {\bibinfo {volume} {93}},\ \bibinfo
  {pages} {134207} (\bibinfo {year} {2016})}\BibitemShut {NoStop}%
\bibitem [{\citenamefont {Berkelbach}\ and\ \citenamefont
  {Reichman}(2010)}]{berkelbach10}%
  \BibitemOpen
  \bibfield  {author} {\bibinfo {author} {\bibfnamefont {T.~C.}\ \bibnamefont
  {Berkelbach}}\ and\ \bibinfo {author} {\bibfnamefont {D.~R.}\ \bibnamefont
  {Reichman}},\ }\href {\doibase 10.1103/PhysRevB.81.224429} {\bibfield
  {journal} {\bibinfo  {journal} {Phys. Rev. B}\ }\textbf {\bibinfo {volume}
  {81}},\ \bibinfo {pages} {224429} (\bibinfo {year} {2010})}\BibitemShut
  {NoStop}%
\bibitem [{\citenamefont {Bari\ifmmode \check{s}\else
  \v{s}\fi{}i\ifmmode~\acute{c}\else \'{c}\fi{}}\ and\ \citenamefont
  {Prelov\ifmmode~\check{s}\else \v{s}\fi{}ek}(2010)}]{barisic10}%
  \BibitemOpen
  \bibfield  {author} {\bibinfo {author} {\bibfnamefont {O.~S.}\ \bibnamefont
  {Bari\ifmmode \check{s}\else \v{s}\fi{}i\ifmmode~\acute{c}\else \'{c}\fi{}}}\
  and\ \bibinfo {author} {\bibfnamefont {P.}~\bibnamefont
  {Prelov\ifmmode~\check{s}\else \v{s}\fi{}ek}},\ }\href {\doibase
  10.1103/PhysRevB.82.161106} {\bibfield  {journal} {\bibinfo  {journal} {Phys.
  Rev. B}\ }\textbf {\bibinfo {volume} {82}},\ \bibinfo {pages} {161106}
  (\bibinfo {year} {2010})}\BibitemShut {NoStop}%
\bibitem [{\citenamefont {Agarwal}\ \emph
  {et~al.}(2015{\natexlab{a}})\citenamefont {Agarwal}, \citenamefont
  {Gopalakrishnan}, \citenamefont {Knap}, \citenamefont {M\"uller},\ and\
  \citenamefont {Demler}}]{agarwal15}%
  \BibitemOpen
  \bibfield  {author} {\bibinfo {author} {\bibfnamefont {K.}~\bibnamefont
  {Agarwal}}, \bibinfo {author} {\bibfnamefont {S.}~\bibnamefont
  {Gopalakrishnan}}, \bibinfo {author} {\bibfnamefont {M.}~\bibnamefont
  {Knap}}, \bibinfo {author} {\bibfnamefont {M.}~\bibnamefont {M\"uller}}, \
  and\ \bibinfo {author} {\bibfnamefont {E.}~\bibnamefont {Demler}},\ }\href
  {\doibase 10.1103/PhysRevLett.114.160401} {\bibfield  {journal} {\bibinfo
  {journal} {Phys. Rev. Lett.}\ }\textbf {\bibinfo {volume} {114}},\ \bibinfo
  {pages} {160401} (\bibinfo {year} {2015}{\natexlab{a}})}\BibitemShut
  {NoStop}%
\bibitem [{\citenamefont {Gopalakrishnan}\ \emph {et~al.}(2015)\citenamefont
  {Gopalakrishnan}, \citenamefont {M\"uller}, \citenamefont {Khemani},
  \citenamefont {Knap}, \citenamefont {Demler},\ and\ \citenamefont
  {Huse}}]{gopal15}%
  \BibitemOpen
  \bibfield  {author} {\bibinfo {author} {\bibfnamefont {S.}~\bibnamefont
  {Gopalakrishnan}}, \bibinfo {author} {\bibfnamefont {M.}~\bibnamefont
  {M\"uller}}, \bibinfo {author} {\bibfnamefont {V.}~\bibnamefont {Khemani}},
  \bibinfo {author} {\bibfnamefont {M.}~\bibnamefont {Knap}}, \bibinfo {author}
  {\bibfnamefont {E.}~\bibnamefont {Demler}}, \ and\ \bibinfo {author}
  {\bibfnamefont {D.~A.}\ \bibnamefont {Huse}},\ }\href {\doibase
  10.1103/PhysRevB.92.104202} {\bibfield  {journal} {\bibinfo  {journal} {Phys.
  Rev. B}\ }\textbf {\bibinfo {volume} {92}},\ \bibinfo {pages} {104202}
  (\bibinfo {year} {2015})}\BibitemShut {NoStop}%
\bibitem [{\citenamefont {Bar~Lev}\ \emph {et~al.}(2015)\citenamefont
  {Bar~Lev}, \citenamefont {Cohen},\ and\ \citenamefont {Reichman}}]{lev15}%
  \BibitemOpen
  \bibfield  {author} {\bibinfo {author} {\bibfnamefont {Y.}~\bibnamefont
  {Bar~Lev}}, \bibinfo {author} {\bibfnamefont {G.}~\bibnamefont {Cohen}}, \
  and\ \bibinfo {author} {\bibfnamefont {D.~R.}\ \bibnamefont {Reichman}},\
  }\href {\doibase 10.1103/PhysRevLett.114.100601} {\bibfield  {journal}
  {\bibinfo  {journal} {Phys. Rev. Lett.}\ }\textbf {\bibinfo {volume} {114}},\
  \bibinfo {pages} {100601} (\bibinfo {year} {2015})}\BibitemShut {NoStop}%
\bibitem [{\citenamefont {Steinigeweg}\ \emph {et~al.}()\citenamefont
  {Steinigeweg}, \citenamefont {Herbrych}, \citenamefont {Pollmann},\ and\
  \citenamefont {Brenig}}]{steinigeweg15}%
  \BibitemOpen
  \bibfield  {author} {\bibinfo {author} {\bibfnamefont {R.}~\bibnamefont
  {Steinigeweg}}, \bibinfo {author} {\bibfnamefont {J.}~\bibnamefont
  {Herbrych}}, \bibinfo {author} {\bibfnamefont {F.}~\bibnamefont {Pollmann}},
  \ and\ \bibinfo {author} {\bibfnamefont {W.}~\bibnamefont {Brenig}},\
  }\href@noop {} {\ }\Eprint {http://arxiv.org/abs/1512.08519}
  {arXiv:1512.08519} \BibitemShut {NoStop}%
\bibitem [{\citenamefont {Bari\ifmmode \check{s}\else
  \v{s}\fi{}i\ifmmode~\acute{c}\else \'{c}\fi{}}\ \emph
  {et~al.}(2016)\citenamefont {Bari\ifmmode \check{s}\else
  \v{s}\fi{}i\ifmmode~\acute{c}\else \'{c}\fi{}}, \citenamefont {Kokalj},
  \citenamefont {Balog},\ and\ \citenamefont {Prelov\ifmmode~\check{s}\else
  \v{s}\fi{}ek}}]{barisic16}%
  \BibitemOpen
  \bibfield  {author} {\bibinfo {author} {\bibfnamefont {O.~S.}\ \bibnamefont
  {Bari\ifmmode \check{s}\else \v{s}\fi{}i\ifmmode~\acute{c}\else \'{c}\fi{}}},
  \bibinfo {author} {\bibfnamefont {J.}~\bibnamefont {Kokalj}}, \bibinfo
  {author} {\bibfnamefont {I.}~\bibnamefont {Balog}}, \ and\ \bibinfo {author}
  {\bibfnamefont {P.}~\bibnamefont {Prelov\ifmmode~\check{s}\else
  \v{s}\fi{}ek}},\ }\href {\doibase 10.1103/PhysRevB.94.045126} {\bibfield
  {journal} {\bibinfo  {journal} {Phys. Rev. B}\ }\textbf {\bibinfo {volume}
  {94}},\ \bibinfo {pages} {045126} (\bibinfo {year} {2016})}\BibitemShut
  {NoStop}%
\bibitem [{\citenamefont {{\v Znidari\v c}}\ \emph {et~al.}(2016)\citenamefont
  {{\v Znidari\v c}}, \citenamefont {Scardicchio},\ and\ \citenamefont
  {Varma}}]{znidaric16}%
  \BibitemOpen
  \bibfield  {author} {\bibinfo {author} {\bibfnamefont {M.}~\bibnamefont {{\v
  Znidari\v c}}}, \bibinfo {author} {\bibfnamefont {A.}~\bibnamefont
  {Scardicchio}}, \ and\ \bibinfo {author} {\bibfnamefont {V.~K.}\ \bibnamefont
  {Varma}},\ }\href {http://dx.doi.org/10.1103/PhysRevLett.117.040601}
  {\bibfield  {journal} {\bibinfo  {journal} {Phys.~Rev.~Lett.}\ }\textbf
  {\bibinfo {volume} {117}},\ \bibinfo {pages} {040601} (\bibinfo {year}
  {2016})}\BibitemShut {NoStop}%
\bibitem [{\citenamefont {\v{Z}nidari\v{c}}\ \emph {et~al.}(2008)\citenamefont
  {\v{Z}nidari\v{c}}, \citenamefont {Prosen},\ and\ \citenamefont
  {Prelov\ifmmode~\check{s}\else \v{s}\fi{}ek}}]{znidaric08}%
  \BibitemOpen
  \bibfield  {author} {\bibinfo {author} {\bibfnamefont {M.}~\bibnamefont
  {\v{Z}nidari\v{c}}}, \bibinfo {author} {\bibfnamefont {T.}~\bibnamefont
  {Prosen}}, \ and\ \bibinfo {author} {\bibfnamefont {P.}~\bibnamefont
  {Prelov\ifmmode~\check{s}\else \v{s}\fi{}ek}},\ }\href {\doibase
  10.1103/PhysRevB.77.064426} {\bibfield  {journal} {\bibinfo  {journal} {Phys.
  Rev. B}\ }\textbf {\bibinfo {volume} {77}},\ \bibinfo {pages} {064426}
  (\bibinfo {year} {2008})}\BibitemShut {NoStop}%
\bibitem [{\citenamefont {Bardarson}\ \emph {et~al.}(2012)\citenamefont
  {Bardarson}, \citenamefont {Pollmann},\ and\ \citenamefont
  {Moore}}]{bardarson12}%
  \BibitemOpen
  \bibfield  {author} {\bibinfo {author} {\bibfnamefont {J.~H.}\ \bibnamefont
  {Bardarson}}, \bibinfo {author} {\bibfnamefont {F.}~\bibnamefont {Pollmann}},
  \ and\ \bibinfo {author} {\bibfnamefont {J.~E.}\ \bibnamefont {Moore}},\
  }\href {\doibase 10.1103/PhysRevLett.109.017202} {\bibfield  {journal}
  {\bibinfo  {journal} {Phys. Rev. Lett.}\ }\textbf {\bibinfo {volume} {109}},\
  \bibinfo {pages} {017202} (\bibinfo {year} {2012})}\BibitemShut {NoStop}%
\bibitem [{\citenamefont {Serbyn}\ \emph {et~al.}(2015)\citenamefont {Serbyn},
  \citenamefont {Papi{\'{c}}},\ and\ \citenamefont {Abanin}}]{serbyn15}%
  \BibitemOpen
  \bibfield  {author} {\bibinfo {author} {\bibfnamefont {M.}~\bibnamefont
  {Serbyn}}, \bibinfo {author} {\bibfnamefont {Z.}~\bibnamefont {Papi{\'{c}}}},
  \ and\ \bibinfo {author} {\bibfnamefont {D.~A.}\ \bibnamefont {Abanin}},\
  }\href {\doibase 10.1103/PhysRevX.5.041047} {\bibfield  {journal} {\bibinfo
  {journal} {Phys. Rev. X}\ }\textbf {\bibinfo {volume} {5}},\ \bibinfo {pages}
  {041047} (\bibinfo {year} {2015})}\BibitemShut {NoStop}%
\bibitem [{\citenamefont {Huse}\ \emph {et~al.}(2014)\citenamefont {Huse},
  \citenamefont {Nandkishore},\ and\ \citenamefont {Oganesyan}}]{huse14}%
  \BibitemOpen
  \bibfield  {author} {\bibinfo {author} {\bibfnamefont {D.~A.}\ \bibnamefont
  {Huse}}, \bibinfo {author} {\bibfnamefont {R.}~\bibnamefont {Nandkishore}}, \
  and\ \bibinfo {author} {\bibfnamefont {V.}~\bibnamefont {Oganesyan}},\ }\href
  {\doibase 10.1103/PhysRevB.90.174202} {\bibfield  {journal} {\bibinfo
  {journal} {Phys. Rev. B}\ }\textbf {\bibinfo {volume} {90}},\ \bibinfo
  {pages} {174202} (\bibinfo {year} {2014})}\BibitemShut {NoStop}%
\bibitem [{\citenamefont {Serbyn}\ \emph {et~al.}(2013)\citenamefont {Serbyn},
  \citenamefont {Papi\ifmmode~\acute{c}\else \'{c}\fi{}},\ and\ \citenamefont
  {Abanin}}]{serbyn13}%
  \BibitemOpen
  \bibfield  {author} {\bibinfo {author} {\bibfnamefont {M.}~\bibnamefont
  {Serbyn}}, \bibinfo {author} {\bibfnamefont {Z.}~\bibnamefont
  {Papi\ifmmode~\acute{c}\else \'{c}\fi{}}}, \ and\ \bibinfo {author}
  {\bibfnamefont {D.~A.}\ \bibnamefont {Abanin}},\ }\href {\doibase
  10.1103/PhysRevLett.111.127201} {\bibfield  {journal} {\bibinfo  {journal}
  {Phys. Rev. Lett.}\ }\textbf {\bibinfo {volume} {111}},\ \bibinfo {pages}
  {127201} (\bibinfo {year} {2013})}\BibitemShut {NoStop}%
\bibitem [{\citenamefont {Ros}\ \emph {et~al.}(2015)\citenamefont {Ros},
  \citenamefont {M{\"u}ller},\ and\ \citenamefont {Scardicchio}}]{Ros:2015}%
  \BibitemOpen
  \bibfield  {author} {\bibinfo {author} {\bibfnamefont {V.}~\bibnamefont
  {Ros}}, \bibinfo {author} {\bibfnamefont {M.}~\bibnamefont {M{\"u}ller}}, \
  and\ \bibinfo {author} {\bibfnamefont {A.}~\bibnamefont {Scardicchio}},\
  }\href {http://www.sciencedirect.com/science/article/pii/S0550321314003836}
  {\bibfield  {journal} {\bibinfo  {journal} {Nuclear Physics B}\ }\textbf
  {\bibinfo {volume} {891}},\ \bibinfo {pages} {420} (\bibinfo {year}
  {2015})}\BibitemShut {NoStop}%
\bibitem [{\citenamefont {Imbrie}(2016)}]{Imbrie:2016}%
  \BibitemOpen
  \bibfield  {author} {\bibinfo {author} {\bibfnamefont {J.~Z.}\ \bibnamefont
  {Imbrie}},\ }\href
  {http://link.springer.com/article/10.1007/s10955-016-1508-x} {\bibfield
  {journal} {\bibinfo  {journal} {J.~Stat.~Phys.}\ }\textbf {\bibinfo {volume}
  {163}},\ \bibinfo {pages} {998} (\bibinfo {year} {2016})}\BibitemShut
  {NoStop}%
\bibitem [{\citenamefont {Monthus}\ and\ \citenamefont
  {Garel}(2010)}]{monthus10}%
  \BibitemOpen
  \bibfield  {author} {\bibinfo {author} {\bibfnamefont {C.}~\bibnamefont
  {Monthus}}\ and\ \bibinfo {author} {\bibfnamefont {T.}~\bibnamefont
  {Garel}},\ }\href {\doibase 10.1103/PhysRevB.81.134202} {\bibfield  {journal}
  {\bibinfo  {journal} {Phys. Rev. B}\ }\textbf {\bibinfo {volume} {81}},\
  \bibinfo {pages} {134202} (\bibinfo {year} {2010})}\BibitemShut {NoStop}%
\bibitem [{\citenamefont {Pal}\ and\ \citenamefont {Huse}(2010)}]{pal10}%
  \BibitemOpen
  \bibfield  {author} {\bibinfo {author} {\bibfnamefont {A.}~\bibnamefont
  {Pal}}\ and\ \bibinfo {author} {\bibfnamefont {D.~A.}\ \bibnamefont {Huse}},\
  }\href {\doibase 10.1103/PhysRevB.82.174411} {\bibfield  {journal} {\bibinfo
  {journal} {Phys. Rev. B}\ }\textbf {\bibinfo {volume} {82}},\ \bibinfo
  {pages} {174411} (\bibinfo {year} {2010})}\BibitemShut {NoStop}%
\bibitem [{\citenamefont {Luitz}\ \emph {et~al.}(2016)\citenamefont {Luitz},
  \citenamefont {Laflorencie},\ and\ \citenamefont {Alet}}]{luitz16}%
  \BibitemOpen
  \bibfield  {author} {\bibinfo {author} {\bibfnamefont {D.~J.}\ \bibnamefont
  {Luitz}}, \bibinfo {author} {\bibfnamefont {N.}~\bibnamefont {Laflorencie}},
  \ and\ \bibinfo {author} {\bibfnamefont {F.}~\bibnamefont {Alet}},\ }\href
  {\doibase 10.1103/PhysRevB.93.060201} {\bibfield  {journal} {\bibinfo
  {journal} {Phys. Rev. B}\ }\textbf {\bibinfo {volume} {93}},\ \bibinfo
  {pages} {060201(R)} (\bibinfo {year} {2016})}\BibitemShut {NoStop}%
\bibitem [{\citenamefont {Mierzejewski}\ \emph {et~al.}()\citenamefont
  {Mierzejewski}, \citenamefont {Herbrych},\ and\ \citenamefont
  {Prelov{\v{s}}ek}}]{mierzejewski16}%
  \BibitemOpen
  \bibfield  {author} {\bibinfo {author} {\bibfnamefont {M.}~\bibnamefont
  {Mierzejewski}}, \bibinfo {author} {\bibfnamefont {J.}~\bibnamefont
  {Herbrych}}, \ and\ \bibinfo {author} {\bibfnamefont {P.}~\bibnamefont
  {Prelov{\v{s}}ek}},\ }\href {http://arxiv.org/abs/1607.04992} {\ }\Eprint
  {http://arxiv.org/abs/1607.04992} {arXiv:1607.04992} \BibitemShut {NoStop}%
\bibitem [{\citenamefont {Prelov{\v{s}}ek}\ and\ \citenamefont
  {Herbrych}()}]{prelovsek16}%
  \BibitemOpen
  \bibfield  {author} {\bibinfo {author} {\bibfnamefont {P.}~\bibnamefont
  {Prelov{\v{s}}ek}}\ and\ \bibinfo {author} {\bibfnamefont {J.}~\bibnamefont
  {Herbrych}},\ }\href {http://arxiv.org/abs/1609.05450} {\ }\Eprint
  {http://arxiv.org/abs/1609.05450} {arXiv:1609.05450} \BibitemShut {NoStop}%
\bibitem [{\citenamefont {Huse}\ \emph {et~al.}(2013)\citenamefont {Huse},
  \citenamefont {Nandkishore}, \citenamefont {Oganesyan}, \citenamefont {Pal},\
  and\ \citenamefont {Sondhi}}]{huse13}%
  \BibitemOpen
  \bibfield  {author} {\bibinfo {author} {\bibfnamefont {D.~A.}\ \bibnamefont
  {Huse}}, \bibinfo {author} {\bibfnamefont {R.}~\bibnamefont {Nandkishore}},
  \bibinfo {author} {\bibfnamefont {V.}~\bibnamefont {Oganesyan}}, \bibinfo
  {author} {\bibfnamefont {A.}~\bibnamefont {Pal}}, \ and\ \bibinfo {author}
  {\bibfnamefont {S.~L.}\ \bibnamefont {Sondhi}},\ }\href {\doibase
  10.1103/PhysRevB.88.014206} {\bibfield  {journal} {\bibinfo  {journal} {Phys.
  Rev. B}\ }\textbf {\bibinfo {volume} {88}},\ \bibinfo {pages} {014206}
  (\bibinfo {year} {2013})}\BibitemShut {NoStop}%
\bibitem [{\citenamefont {Chandran}\ \emph {et~al.}(2014)\citenamefont
  {Chandran}, \citenamefont {Khemani}, \citenamefont {Laumann},\ and\
  \citenamefont {Sondhi}}]{Chandran:2014}%
  \BibitemOpen
  \bibfield  {author} {\bibinfo {author} {\bibfnamefont {A.}~\bibnamefont
  {Chandran}}, \bibinfo {author} {\bibfnamefont {V.}~\bibnamefont {Khemani}},
  \bibinfo {author} {\bibfnamefont {C.~R.}\ \bibnamefont {Laumann}}, \ and\
  \bibinfo {author} {\bibfnamefont {S.~L.}\ \bibnamefont {Sondhi}},\ }\href
  {http://journals.aps.org/prb/abstract/10.1103/PhysRevB.89.144201} {\bibfield
  {journal} {\bibinfo  {journal} {Phys. Rev. B}\ }\textbf {\bibinfo {volume}
  {89}},\ \bibinfo {pages} {144201} (\bibinfo {year} {2014})}\BibitemShut
  {NoStop}%
\bibitem [{\citenamefont {Nandkishore}\ and\ \citenamefont
  {Huse}(2015)}]{Nandkishore:2015}%
  \BibitemOpen
  \bibfield  {author} {\bibinfo {author} {\bibfnamefont {R.}~\bibnamefont
  {Nandkishore}}\ and\ \bibinfo {author} {\bibfnamefont {D.~A.}\ \bibnamefont
  {Huse}},\ }\href
  {http://www.annualreviews.org/doi/10.1146/annurev-conmatphys-031214-014726}
  {\bibfield  {journal} {\bibinfo  {journal} {Annu. Rev. Condens. Matter
  Phys.}\ }\textbf {\bibinfo {volume} {6}},\ \bibinfo {pages} {15} (\bibinfo
  {year} {2015})}\BibitemShut {NoStop}%
\bibitem [{\citenamefont {Altman}\ and\ \citenamefont
  {Vosk}(2015)}]{Altman:2015}%
  \BibitemOpen
  \bibfield  {author} {\bibinfo {author} {\bibfnamefont {E.}~\bibnamefont
  {Altman}}\ and\ \bibinfo {author} {\bibfnamefont {R.}~\bibnamefont {Vosk}},\
  }\href
  {http://www.annualreviews.org/doi/abs/10.1146/annurev-conmatphys-031214-014701}
  {\bibfield  {journal} {\bibinfo  {journal} {Annu. Rev. Condens. Matter
  Phys.}\ }\textbf {\bibinfo {volume} {6}},\ \bibinfo {pages} {383} (\bibinfo
  {year} {2015})}\BibitemShut {NoStop}%
\bibitem [{\citenamefont {Schreiber}\ \emph {et~al.}(2015)\citenamefont
  {Schreiber}, \citenamefont {Hodgman}, \citenamefont {Bordia}, \citenamefont
  {L{\"{u}}schen}, \citenamefont {Fischer}, \citenamefont {Vosk}, \citenamefont
  {Altman}, \citenamefont {Schneider},\ and\ \citenamefont
  {Bloch}}]{schreiber15}%
  \BibitemOpen
  \bibfield  {author} {\bibinfo {author} {\bibfnamefont {M.}~\bibnamefont
  {Schreiber}}, \bibinfo {author} {\bibfnamefont {S.~S.}\ \bibnamefont
  {Hodgman}}, \bibinfo {author} {\bibfnamefont {P.}~\bibnamefont {Bordia}},
  \bibinfo {author} {\bibfnamefont {H.~P.}\ \bibnamefont {L{\"{u}}schen}},
  \bibinfo {author} {\bibfnamefont {M.~H.}\ \bibnamefont {Fischer}}, \bibinfo
  {author} {\bibfnamefont {R.}~\bibnamefont {Vosk}}, \bibinfo {author}
  {\bibfnamefont {E.}~\bibnamefont {Altman}}, \bibinfo {author} {\bibfnamefont
  {U.}~\bibnamefont {Schneider}}, \ and\ \bibinfo {author} {\bibfnamefont
  {I.}~\bibnamefont {Bloch}},\ }\href {\doibase 10.1126/science.aaa7432}
  {\bibfield  {journal} {\bibinfo  {journal} {Science}\ }\textbf {\bibinfo
  {volume} {349}},\ \bibinfo {pages} {842} (\bibinfo {year}
  {2015})}\BibitemShut {NoStop}%
\bibitem [{\citenamefont {Kondov}\ \emph {et~al.}(2015)\citenamefont {Kondov},
  \citenamefont {McGehee}, \citenamefont {Xu},\ and\ \citenamefont
  {DeMarco}}]{kondov15}%
  \BibitemOpen
  \bibfield  {author} {\bibinfo {author} {\bibfnamefont {S.~S.}\ \bibnamefont
  {Kondov}}, \bibinfo {author} {\bibfnamefont {W.~R.}\ \bibnamefont {McGehee}},
  \bibinfo {author} {\bibfnamefont {W.}~\bibnamefont {Xu}}, \ and\ \bibinfo
  {author} {\bibfnamefont {B.}~\bibnamefont {DeMarco}},\ }\href {\doibase
  10.1103/PhysRevLett.114.083002} {\bibfield  {journal} {\bibinfo  {journal}
  {Phys. Rev. Lett.}\ }\textbf {\bibinfo {volume} {114}},\ \bibinfo {pages}
  {083002} (\bibinfo {year} {2015})}\BibitemShut {NoStop}%
\bibitem [{\citenamefont {Bordia}\ \emph {et~al.}(2016)\citenamefont {Bordia},
  \citenamefont {L{\"{u}}schen}, \citenamefont {Hodgman}, \citenamefont
  {Schreiber}, \citenamefont {Bloch},\ and\ \citenamefont
  {Schneider}}]{bordia16}%
  \BibitemOpen
  \bibfield  {author} {\bibinfo {author} {\bibfnamefont {P.}~\bibnamefont
  {Bordia}}, \bibinfo {author} {\bibfnamefont {H.~P.}\ \bibnamefont
  {L{\"{u}}schen}}, \bibinfo {author} {\bibfnamefont {S.~S.}\ \bibnamefont
  {Hodgman}}, \bibinfo {author} {\bibfnamefont {M.}~\bibnamefont {Schreiber}},
  \bibinfo {author} {\bibfnamefont {I.}~\bibnamefont {Bloch}}, \ and\ \bibinfo
  {author} {\bibfnamefont {U.}~\bibnamefont {Schneider}},\ }\href {\doibase
  10.1103/PhysRevLett.116.140401} {\bibfield  {journal} {\bibinfo  {journal}
  {Phys. Rev. Lett.}\ }\textbf {\bibinfo {volume} {116}},\ \bibinfo {pages}
  {140401} (\bibinfo {year} {2016})}\BibitemShut {NoStop}%
\bibitem [{\citenamefont {Choi}\ \emph {et~al.}(2016)\citenamefont {Choi},
  \citenamefont {Hild}, \citenamefont {Zeiher}, \citenamefont {Schau{\ss}},
  \citenamefont {Rubio-Abadal}, \citenamefont {Yefsah}, \citenamefont
  {Khemani}, \citenamefont {Huse}, \citenamefont {Bloch},\ and\ \citenamefont
  {Gross}}]{choi16}%
  \BibitemOpen
  \bibfield  {author} {\bibinfo {author} {\bibfnamefont {J.-Y.}\ \bibnamefont
  {Choi}}, \bibinfo {author} {\bibfnamefont {S.}~\bibnamefont {Hild}}, \bibinfo
  {author} {\bibfnamefont {J.}~\bibnamefont {Zeiher}}, \bibinfo {author}
  {\bibfnamefont {P.}~\bibnamefont {Schau{\ss}}}, \bibinfo {author}
  {\bibfnamefont {A.}~\bibnamefont {Rubio-Abadal}}, \bibinfo {author}
  {\bibfnamefont {T.}~\bibnamefont {Yefsah}}, \bibinfo {author} {\bibfnamefont
  {V.}~\bibnamefont {Khemani}}, \bibinfo {author} {\bibfnamefont {D.~A.}\
  \bibnamefont {Huse}}, \bibinfo {author} {\bibfnamefont {I.}~\bibnamefont
  {Bloch}}, \ and\ \bibinfo {author} {\bibfnamefont {C.}~\bibnamefont
  {Gross}},\ }\href {\doibase 10.1126/science.aaf8834} {\bibfield  {journal}
  {\bibinfo  {journal} {Science}\ }\textbf {\bibinfo {volume} {352}},\ \bibinfo
  {pages} {1547} (\bibinfo {year} {2016})}\BibitemShut {NoStop}%
\bibitem [{\citenamefont {Smith}\ \emph {et~al.}(2016)\citenamefont {Smith},
  \citenamefont {Lee}, \citenamefont {Richerme}, \citenamefont {Neyenhuis},
  \citenamefont {Hess}, \citenamefont {Hauke}, \citenamefont {Heyl},
  \citenamefont {Huse},\ and\ \citenamefont {Monroe}}]{Maryland}%
  \BibitemOpen
  \bibfield  {author} {\bibinfo {author} {\bibfnamefont {J.}~\bibnamefont
  {Smith}}, \bibinfo {author} {\bibfnamefont {A.}~\bibnamefont {Lee}}, \bibinfo
  {author} {\bibfnamefont {P.}~\bibnamefont {Richerme}}, \bibinfo {author}
  {\bibfnamefont {B.}~\bibnamefont {Neyenhuis}}, \bibinfo {author}
  {\bibfnamefont {P.~W.}\ \bibnamefont {Hess}}, \bibinfo {author}
  {\bibfnamefont {P.}~\bibnamefont {Hauke}}, \bibinfo {author} {\bibfnamefont
  {M.}~\bibnamefont {Heyl}}, \bibinfo {author} {\bibfnamefont {D.~A.}\
  \bibnamefont {Huse}}, \ and\ \bibinfo {author} {\bibfnamefont
  {C.}~\bibnamefont {Monroe}},\ }\href {\doibase dx.doi.org/10.1038/nphys3783}
  {\bibfield  {journal} {\bibinfo  {journal} {Nat. Phys.}\ }\textbf {\bibinfo
  {volume} {advance online publication}} (\bibinfo {year} {2016}),\
  dx.doi.org/10.1038/nphys3783}\BibitemShut {NoStop}%
\bibitem [{\citenamefont {Mondaini}\ and\ \citenamefont
  {Rigol}(2015)}]{mondaini15}%
  \BibitemOpen
  \bibfield  {author} {\bibinfo {author} {\bibfnamefont {R.}~\bibnamefont
  {Mondaini}}\ and\ \bibinfo {author} {\bibfnamefont {M.}~\bibnamefont
  {Rigol}},\ }\href {\doibase 10.1103/PhysRevA.92.041601} {\bibfield  {journal}
  {\bibinfo  {journal} {Phys. Rev. A}\ }\textbf {\bibinfo {volume} {92}},\
  \bibinfo {pages} {041601(R)} (\bibinfo {year} {2015})}\BibitemShut {NoStop}%
\bibitem [{\citenamefont {{Bar Lev}}\ and\ \citenamefont
  {Reichman}(2016)}]{barlev16}%
  \BibitemOpen
  \bibfield  {author} {\bibinfo {author} {\bibfnamefont {Y.}~\bibnamefont {{Bar
  Lev}}}\ and\ \bibinfo {author} {\bibfnamefont {D.~R.}\ \bibnamefont
  {Reichman}},\ }\href {\doibase 10.1209/0295-5075/113/46001} {\bibfield
  {journal} {\bibinfo  {journal} {EPL (Europhysics Letters)}\ }\textbf
  {\bibinfo {volume} {113}},\ \bibinfo {pages} {46001} (\bibinfo {year}
  {2016})}\BibitemShut {NoStop}%
\bibitem [{sup()}]{suppl}%
  \BibitemOpen
  \href@noop {} {}\bibinfo {note} {See Supplemental Material, which includes
  Refs.~\cite{schreiber15,bordia16,mondaini15,shastry86,*prosen12,znidaric13},
  for data for different $L$, quarter-filling initial states, Aubry-Andre
  quasiperiodic disorder, and demonstration of global spin
  relaxation.}\BibitemShut {Stop}%
\bibitem [{\citenamefont {Long}\ \emph {et~al.}(2003)\citenamefont {Long},
  \citenamefont {Prelov\ifmmode~\check{s}\else \v{s}\fi{}ek}, \citenamefont
  {El~Shawish}, \citenamefont {Karadamoglou},\ and\ \citenamefont
  {Zotos}}]{long03}%
  \BibitemOpen
  \bibfield  {author} {\bibinfo {author} {\bibfnamefont {M.~W.}\ \bibnamefont
  {Long}}, \bibinfo {author} {\bibfnamefont {P.}~\bibnamefont
  {Prelov\ifmmode~\check{s}\else \v{s}\fi{}ek}}, \bibinfo {author}
  {\bibfnamefont {S.}~\bibnamefont {El~Shawish}}, \bibinfo {author}
  {\bibfnamefont {J.}~\bibnamefont {Karadamoglou}}, \ and\ \bibinfo {author}
  {\bibfnamefont {X.}~\bibnamefont {Zotos}},\ }\href {\doibase
  10.1103/PhysRevB.68.235106} {\bibfield  {journal} {\bibinfo  {journal} {Phys.
  Rev. B}\ }\textbf {\bibinfo {volume} {68}},\ \bibinfo {pages} {235106}
  (\bibinfo {year} {2003})}\BibitemShut {NoStop}%
\bibitem [{\citenamefont {Prelov\v{s}ek}\ and\ \citenamefont
  {Bon\v{c}a}(2013)}]{prelovsek13}%
  \BibitemOpen
  \bibfield  {author} {\bibinfo {author} {\bibfnamefont {P.}~\bibnamefont
  {Prelov\v{s}ek}}\ and\ \bibinfo {author} {\bibfnamefont {J.}~\bibnamefont
  {Bon\v{c}a}},\ }\bibfield  {booktitle} {\emph {\bibinfo {booktitle} {Strongly
  Correlated Systems - Numerical Methods}},\ }\href@noop {} {\  (\bibinfo
  {year} {2013})}\BibitemShut {NoStop}%
\bibitem [{\citenamefont {Karahalios}\ \emph {et~al.}(2009)\citenamefont
  {Karahalios}, \citenamefont {Metavitsiadis}, \citenamefont {Zotos},
  \citenamefont {Gorczyca},\ and\ \citenamefont {Prelov\ifmmode~\check{s}\else
  \v{s}\fi{}ek}}]{karahalios09}%
  \BibitemOpen
  \bibfield  {author} {\bibinfo {author} {\bibfnamefont {A.}~\bibnamefont
  {Karahalios}}, \bibinfo {author} {\bibfnamefont {A.}~\bibnamefont
  {Metavitsiadis}}, \bibinfo {author} {\bibfnamefont {X.}~\bibnamefont
  {Zotos}}, \bibinfo {author} {\bibfnamefont {A.}~\bibnamefont {Gorczyca}}, \
  and\ \bibinfo {author} {\bibfnamefont {P.}~\bibnamefont
  {Prelov\ifmmode~\check{s}\else \v{s}\fi{}ek}},\ }\href {\doibase
  10.1103/PhysRevB.79.024425} {\bibfield  {journal} {\bibinfo  {journal} {Phys.
  Rev. B}\ }\textbf {\bibinfo {volume} {79}},\ \bibinfo {pages} {024425}
  (\bibinfo {year} {2009})}\BibitemShut {NoStop}%
\bibitem [{\citenamefont {Agarwal}\ \emph
  {et~al.}(2015{\natexlab{b}})\citenamefont {Agarwal}, \citenamefont {Demler},\
  and\ \citenamefont {Martin}}]{agarwal115}%
  \BibitemOpen
  \bibfield  {author} {\bibinfo {author} {\bibfnamefont {K.}~\bibnamefont
  {Agarwal}}, \bibinfo {author} {\bibfnamefont {E.}~\bibnamefont {Demler}}, \
  and\ \bibinfo {author} {\bibfnamefont {I.}~\bibnamefont {Martin}},\ }\href
  {\doibase 10.1103/PhysRevB.92.184203} {\bibfield  {journal} {\bibinfo
  {journal} {Physical Review B - Condensed Matter and Materials Physics}\
  }\textbf {\bibinfo {volume} {92}},\ \bibinfo {pages} {1} (\bibinfo {year}
  {2015}{\natexlab{b}})}\BibitemShut {NoStop}%
\bibitem [{\citenamefont {Ogata}\ and\ \citenamefont {Shiba}(1990)}]{ogata90}%
  \BibitemOpen
  \bibfield  {author} {\bibinfo {author} {\bibfnamefont {M.}~\bibnamefont
  {Ogata}}\ and\ \bibinfo {author} {\bibfnamefont {H.}~\bibnamefont {Shiba}},\
  }\href {\doibase 10.1103/PhysRevB.41.2326} {\bibfield  {journal} {\bibinfo
  {journal} {Phys. Rev. B}\ }\textbf {\bibinfo {volume} {41}},\ \bibinfo
  {pages} {2326} (\bibinfo {year} {1990})}\BibitemShut {NoStop}%
\bibitem [{\citenamefont {Parameswaran}\ and\ \citenamefont
  {Gopalakrishnan}(2016)}]{parameswaran16}%
  \BibitemOpen
  \bibfield  {author} {\bibinfo {author} {\bibfnamefont {S.~A.}\ \bibnamefont
  {Parameswaran}}\ and\ \bibinfo {author} {\bibfnamefont {S.}~\bibnamefont
  {Gopalakrishnan}},\ }\href {http://arxiv.org/abs/1603.08933} {\  (\bibinfo
  {year} {2016})},\ \Eprint {http://arxiv.org/abs/1603.08933}
  {arXiv:1603.08933} \BibitemShut {NoStop}%
\bibitem [{\citenamefont {Vasseur}\ \emph {et~al.}(2015)\citenamefont
  {Vasseur}, \citenamefont {Potter},\ and\ \citenamefont
  {Parameswaran}}]{vasseur15}%
  \BibitemOpen
  \bibfield  {author} {\bibinfo {author} {\bibfnamefont {R.}~\bibnamefont
  {Vasseur}}, \bibinfo {author} {\bibfnamefont {A.~C.}\ \bibnamefont {Potter}},
  \ and\ \bibinfo {author} {\bibfnamefont {S.}~\bibnamefont {Parameswaran}},\
  }\href {\doibase 10.1103/PhysRevLett.114.217201} {\bibfield  {journal}
  {\bibinfo  {journal} {Phys. Rev. Lett.}\ }\textbf {\bibinfo {volume} {114}},\
  \bibinfo {pages} {217201} (\bibinfo {year} {2015})}\BibitemShut {NoStop}%
\bibitem [{\citenamefont {Potter}\ and\ \citenamefont {Vasseur}()}]{potter16}%
  \BibitemOpen
  \bibfield  {author} {\bibinfo {author} {\bibfnamefont {A.~C.}\ \bibnamefont
  {Potter}}\ and\ \bibinfo {author} {\bibfnamefont {R.}~\bibnamefont
  {Vasseur}},\ }\href {http://arxiv.org/abs/1605.03601} {\ }\Eprint
  {http://arxiv.org/abs/1605.03601} {arXiv:1605.03601} \BibitemShut {NoStop}%
\bibitem [{\citenamefont {Shastry}(1986)}]{shastry86}%
  \BibitemOpen
  \bibfield  {author} {\bibinfo {author} {\bibfnamefont {B.~S.}\ \bibnamefont
  {Shastry}},\ }\href {\doibase 10.1103/PhysRevLett.56.1529} {\bibfield
  {journal} {\bibinfo  {journal} {Phys. Rev. Lett.}\ }\textbf {\bibinfo
  {volume} {56}},\ \bibinfo {pages} {1529} (\bibinfo {year}
  {1986})}\BibitemShut {NoStop}%
\bibitem [{\citenamefont {Prosen}\ and\ \citenamefont {{\v Znidari\v
  c}}(2012)}]{prosen12}%
  \BibitemOpen
  \bibfield  {author} {\bibinfo {author} {\bibfnamefont {T.}~\bibnamefont
  {Prosen}}\ and\ \bibinfo {author} {\bibfnamefont {M.}~\bibnamefont {{\v
  Znidari\v c}}},\ }\href {\doibase 10.1103/PhysRevB.86.125118} {\bibfield
  {journal} {\bibinfo  {journal} {Phys. Rev. B}\ }\textbf {\bibinfo {volume}
  {86}},\ \bibinfo {pages} {125118} (\bibinfo {year} {2012})}\BibitemShut
  {NoStop}%
\bibitem [{\citenamefont {\ifmmode \check{Z}\else
  \v{Z}\fi{}nidari\ifmmode~\check{c}\else \v{c}\fi{}}(2013)}]{znidaric13}%
  \BibitemOpen
  \bibfield  {author} {\bibinfo {author} {\bibfnamefont {M.}~\bibnamefont
  {\ifmmode \check{Z}\else \v{Z}\fi{}nidari\ifmmode~\check{c}\else
  \v{c}\fi{}}},\ }\href {\doibase 10.1103/PhysRevLett.110.070602} {\bibfield
  {journal} {\bibinfo  {journal} {Phys. Rev. Lett.}\ }\textbf {\bibinfo
  {volume} {110}},\ \bibinfo {pages} {070602} (\bibinfo {year}
  {2013})}\BibitemShut {NoStop}%
\end{thebibliography}%



\section*{Supplementary material}

# Supplemental material for "Absence of full many-body localization in disordered Hubbard chain"

P. Prelovšek[1,2], O. S. Barišić[3], and M. Žnidarič[2]

[1] Jožef Stefan Institute, SI-1000 Ljubljana, Slovenia
[2] Faculty of Mathematics and Physics, University of Ljubljana, SI-1000 Ljubljana, Slovenia and
[3] Institute of Physics, Zagreb, Croatia

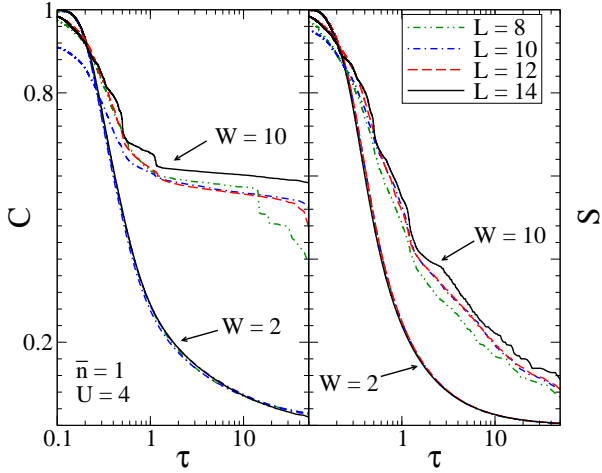

Figure 1. (Color online) Charge and spin imbalance functions, $C(\tau)$ and $S(\tau)$, respectively, obtained using MCLM for different system sizes $L$, all for $U = 4$ and half-filling $\bar{n} = 1$.

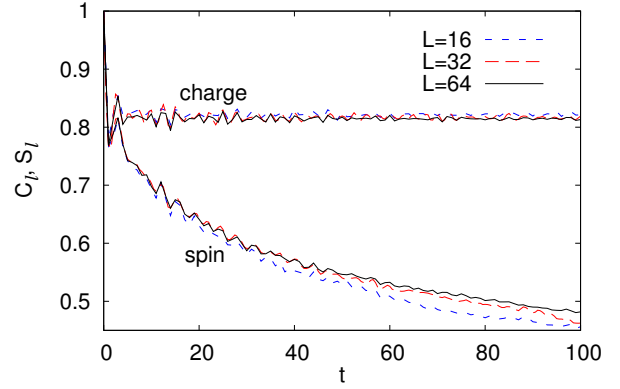

Figure 2. (Color online) Local charge and spin correlations, $C_l(t)$ and $S_l(t)$, obtained by t-DMRG for $U = 1$ and $W = 16$ at half-filling $\bar{n} = 1$ for different system sizes $L$. For $L \geq 32$ results are size-independent.

### Size dependence

Results for the imbalance correlations $C(\tau)$ and $S(\tau)$ have been obtained using MCLM on systems with different length $L$. In order to show that our conclusions do not depend on size restrictions, we present in Fig. 1 results for sizes $L = 8 - 14$, for fixed $U = 4$ and half-filling $\bar{n} = 1$. It is evident that in the ergodic regime, $W = 2$, results for all systems are indistinguishable. On the other hand, beyond the threshold at $W = 10$ results are more delicate. There is no decay in $C(\tau)$ up to $\tau < \tau_m$, but $\tau_m$ depends on $L$ at small sizes (e.g. at $L = 8$) $\tau_m \sim 30$ is already limited by the level spacings and corresponding small $N_L$. For larger sizes $L > 12$ $\tau_m \sim 50$ is also limited, but by reachable $N_L = 10.000$. On the other hand, small sampling $N_s \sim 10$ also influences results for at large $W$, in particular visible in $C(\tau)$.

For results obtained for local charge and spin autocorrelation function, and shown in the main text for $L = 64$, we also here demonstrate that at $L = 64$ the thermodynamic limit is already achieved. In Fig. 2 we show results for $L = 16, 32$ and $64$, demonstrating that for $L \geq 32$ the results are essentially independent of the system size.

### Quarter-filling

We have mainly examined and presented the case of half-filling $\bar{n} = 1$. While this filling might be special for a pure Hubbard model at $T \to 0$, it is expected to be quite generic in the case of considered limit $T \to \infty$ and at large disorder $W$. In order to check this, we consider numerically also the quarter-filled system with $\bar{n} = 1/2$, which is the case realized in cold-atom experiments [1, 2] as well as in previous numerical studies [3]. Results in Fig. 3 indeed confirm that there is no qualitative (and moreover nearly no quantitiative) difference to the $\bar{n} = 1$ results in Fig. 1. While charge imbalance $C(\tau)$ saturates (or shows very slow decay) for $W > 4$, spin correlations $S(\tau)$ appear to decay to zero for all considered $W$, nevertheless still with a change of characteristic time dependence at $W \sim 4$.

### Mapping to a ladder

Using the Jordan-Wigner transformation the 1D Hubbard model can be written as a ladder (upto a constant and a boundary term whose exact form depends on a specific boundary condition used), for details see e.g. Ref. [4],

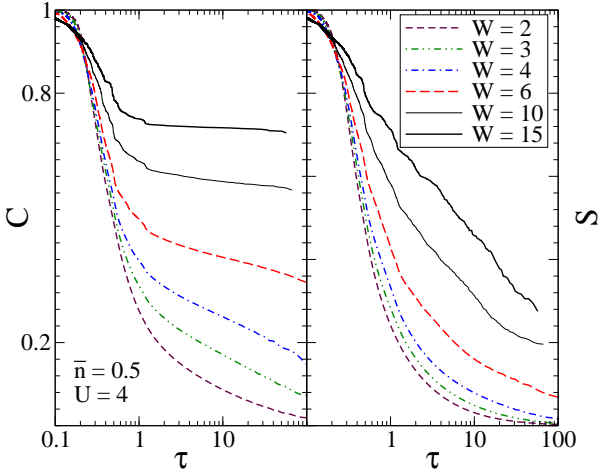

Figure 3. (Color online) $C(\tau)$ and $S(\tau)$, respectively, for $U = 4$ and quarter-filling $\bar{n} = 1/2$, obtained via MCLM within the system size $L = 16$.

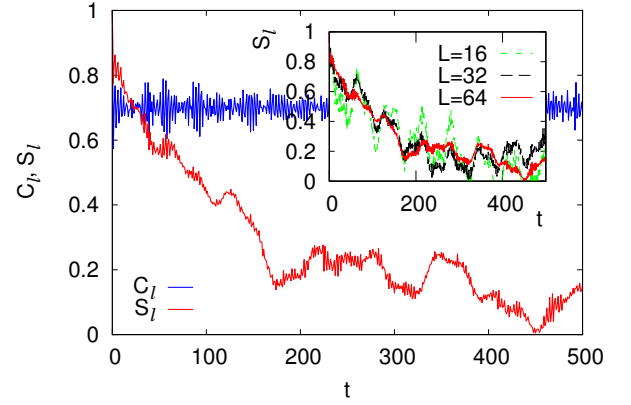

Figure 4. (Color online) Charge and spin autocorrelation function for $U = 1$ and qusiperiodic disorder (2) of strength $W = 8$. Main plot is here for $L = 64$, and, due to lower entanglement, large times can be reached. The inset shows the decay of spin correlation function for different system sizes.

$$H = -\frac{t_0}{2}\sum_{j=1}^{L-1}(\sigma_j^x\sigma_{j+1}^x + \sigma_j^y\sigma_{j+1}^y + \tau_j^x\tau_{j+1}^x + \tau_j^y\tau_{j+1}^y) +$$
$$+\frac{U}{4}\sum_{j=1}^{L}(\sigma_j^z+1)(\tau_j^z+1) + \sum_{j=1}^{L}\frac{\varepsilon_j}{2}(\sigma_j^z+\tau_j^z), \quad (1)$$

where $\sigma_j^\alpha$ and $\tau_j^\alpha$ are two sets of Pauli matrices acting on site $j$ and describing spin-up and spin-down fermions, respectively. We see that the ladder has a "kinetic" XX coupling along legs and an ZZ interaction along rungs. Fermion densities are expressed as $n_{j\uparrow} = (\sigma_j^z+1)/2$ and $n_{j\downarrow} = (\tau_j^z+1)/2$. In such ladder there are exponentially large (but of subleading size in the thermodynamic limit) invariant subspaces possesing ballistic transport, irrespective of the presence and strength of disorder [5]. In our t-DMRG simulations we have simulated the above ladder system using open boundary conditions.

### Aubry-Andre type of disorder

While in the main text we used random independent (charge potential) disorder at each site, it is clear from our symmetry argument that similar decay of spin degrees of freedom happens also for a quasi-periodic disorder, e.g., of the Aubry-Andre type realized in experiments [1, 2]. To numerically demonstrate that this is indeed the case we use the disorder part of $H$ of form $\sum_j \epsilon_j(n_{j\uparrow}+n_{j\downarrow})$, where

$$\epsilon_j = W\cos(2\pi\zeta j), \quad (2)$$

with $\zeta = 0.721$, and used a quarter-filling initial condition $|\uparrow, 0, \downarrow, 0, \uparrow, \ldots\rangle$, which has $n_\uparrow = n_\downarrow = 1/4$.

In Fig. 4 we show results of numerical simulation using t-DMRG method. Again, spin does decay, regardless of the disorder strength. Note that in this case there are larger oscillations/fluctuations simply because there is no averaging – the results are for a single initial state and one quasiperiodic disorder realization (2). For $W = 4$ (data not snown) similar results are obtained, charge is nonergodic while spin decays, though in a shorter time than for $W = 8$.

### Slow global spin relaxation

In a typical initial product state from the half-filling sector there is on average no global charge or spin imbalance and so relaxation of spin happens predominantly locally. This is also a reason why the dependence of $S(\tau)$ and $S_l(t)$ becomes independent of $L$ for large $L$. One can ask if and how the spin, which is apparently not localized, is transported globally on a length-scale $L$ of the whole system. To study global relaxation of spin we take a quarter-filling initial state

$$|\psi(0)\rangle = |\uparrow, 0, \uparrow, \ldots 0, \uparrow, 0, \downarrow, \ldots, 0, \downarrow\rangle, \quad (3)$$

which has a nonzero difference of magnetization in the left and right halves,

$$M(t) = \frac{2}{L}\langle\psi(t)|\left[\sum_{j=1}^{L/2}m_j - \sum_{j=L/2+1}^{L}m_j\right]|\psi(t)\rangle, \quad (4)$$

with the initial value $M(0) = 1$, and the expected long-time value $M(\infty) = 0$ if the spin relaxes globally. In Fig. 5 we show results for $M(t)$ and three different system sizes. As one can see, the suggested scaling variable indicates a global and ergodic transport of spin, which might be of subdiffusive nature.

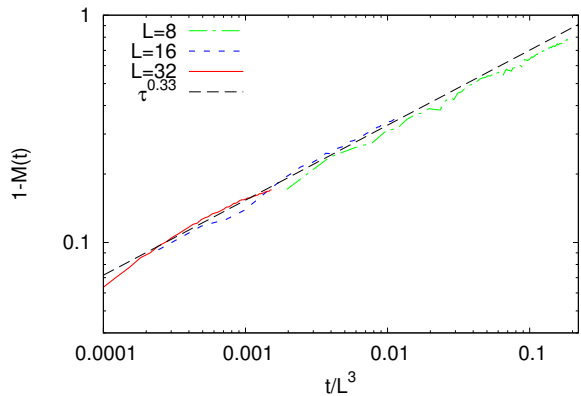

Figure 5. (Color online) Relaxation of global spin for the state with initial difference in left/right magnetization, Eq.(3). For larger systems relaxation of spin gets slower, with the scaling variable seemingly being $\tau = t/L^3$, implying global (but possibly subdiffusive) transport $\Delta x \sim t^{1/3}$. Random charge disorder is of strength $W = 4$ and $U = 1$.

[1] M. Schreiber, S. S. Hodgman, P. Bordia, H. P. Lüschen, M. H. Fischer, R. Vosk, E. Altman, U. Schneider, and I. Bloch, Science **349**, 842 (2015).
[2] P. Bordia, H. P. Lüschen, S. S. Hodgman, M. Schreiber, I. Bloch, and U. Schneider, Phys. Rev. Lett. **116**, 140401 (2016).
[3] R. Mondaini and M. Rigol, Phys. Rev. A **92**, 041601(R) (2015).
[4] B. S. Shastry, Phys. Rev. Lett. **56**, 1529 (1986); T. Prosen and M. Žnidarič, Phys. Rev. B **86**, 125118 (2012).
[5] M. Žnidarič, Phys. Rev. Lett. **110**, 070602 (2013).



\end{document}